\documentclass{article}
\usepackage{arxiv}

\usepackage[utf8]{inputenc} 
\usepackage[T1]{fontenc}    
\usepackage{hyperref}       
\usepackage{url}            
\usepackage{booktabs}       
\usepackage{amsfonts}       
\usepackage{nicefrac}       
\usepackage{graphicx}
\usepackage{natbib}
\usepackage{doi}

\usepackage{amsmath}
\usepackage{comment}

\title{Earthquake Detection Using Benford's Law}

\author{ \href{https://orcid.org/0009-0009-8622-397X}{\includegraphics[scale=0.06]{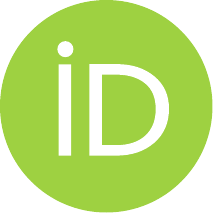}\hspace{1mm}
    Bivas Das} \\
	Department of Earth and Climate Science\\
	Indian Institute of Science Education and Research Pune\\
	Pune 411008, India \\
    \emph{Currently at}  \\
    Observatoire de la C\^ote d'Azur, Universit\'e C\^ote d'Azur \\
    CNRS, IRD, G\'eoazur \\
    Sophia Antipolis, France\\
    \texttt{bivas.das@cnrs.fr} \\
	\And
	\href{https://orcid.org/0000-0002-8249-9399}{\includegraphics[scale=0.06]{orcid.pdf}\hspace{1mm}
    Arjun Datta} \\
	Department of Earth and Climate Science\\
	Indian Institute of Science Education and Research Pune\\
	Pune 411008, India \\
	\texttt{arjundatta@iiserpune.ac.in} \\
    \And
    \href{https://orcid.org/0000-0002-0557-2839}{\includegraphics[scale=0.06]{orcid.pdf}\hspace{1mm}
    Abhijit Ghosh} \\
    Department of Earth and Planetary Sciences \\
    University of California, Riverside \\
    CA, USA \\
    \texttt{aghosh@ucr.edu} \\
    \And
    Visrutha Chalakkatta \\
    Department of Earth and Climate Science \\
    Indian Institute of Science Education and Research Pune \\
    Pune 411008, India \\
}

\begin{document}
\maketitle

\begin{abstract}
    Reliable detection of local earthquakes and accurate identification of P-wave onsets are fundamental tasks in seismology, yet many existing methods to accomplish them require extensive parameter tuning or large training datasets. In this study, we investigate the applicability of Benford’s Law—a logarithmic distribution governing the occurrence of leading digits in naturally occurring data—as a statistical framework for local earthquake detection. We analyze continuous seismic waveform data from two contrasting tectonic environments in the Indian subcontinent: the intraplate Deccan Volcanic Province and the actively deforming Himalayan region. Using a sliding-window approach, we quantify the temporal conformity of first-digit distributions of seismic amplitudes to Benford’s Law and apply an adaptive normalization scheme to account for station-specific noise. Our results show that local earthquake waveforms consistently conform to Benford’s Law during the onset of seismic energy, while pre-event noise does not. Benford-derived statistical anomalies align closely with theoretical P-wave arrival times, with detection performance primarily controlled by window length. These results establish Benford’s Law as a computationally inexpensive, parameter-light, and training-free tool for local earthquake detection.
\end{abstract}




\section{Introduction}
\label{sec:intro}
Seismic event detection and phase identification form the basis of earthquake monitoring, seismic hazard assessment, and imaging of the interior of the Earth. Over the past decades, a wide spectrum of detection approaches has been developed, spanning classical signal-processing techniques, waveform-similarity methods, array-based detectors, and, more recently, machine learning based phase pickers. Among classical approaches, short-term average to long-term average (STA/LTA) algorithms \citep{Allen1978automatic} remain widely used due to their simplicity and computational efficiency, but their performance is often sensitive to parameter tuning and noise conditions. More advanced approaches, such as matched-filter techniques, exploit waveform similarity to detect repeating or low-magnitude events with high sensitivity. Although highly effective for detecting similar or clustered events, these methods require representative templates and become computationally demanding for large continuous datasets. Their applicability is therefore limited in settings where prior templates are unavailable or waveform variability is high. Array-based methods leverage spatial coherence across multiple stations to enhance detection reliability but depend on network geometry and data availability. In parallel, ML and deep learning–based pickers \citep[eg.,][]{zhou2021an,zhou2025AI-PAL,suarez2025picking,li2024saipy} have demonstrated remarkable performance in phase identification, yet typically require large, labeled training datasets and significant computational resources, and may face challenges in generalization across regions, instruments, and noise environments. 

Despite this diversity of approaches, there is a need for detection frameworks that are computationally efficient, minimally parameterized, and independent of prior training or template information. This is particularly true of long-duration deployments, noisy environments, and emerging applications such as planetary seismology or remote terrestrial monitoring, where data availability and prior information may be limited. In such  cases, methods that rely on intrinsic statistical properties of seismic waveforms, rather than explicit amplitude thresholds or learned representations, offer a promising detection alternative. One such statistics-based method is provided by Benford’s Law \citep{benford1938law}, which describes the expected distribution of leading digits in naturally occurring datasets such as physical constants and tabulated scientific quantities \citep{newcomb1881note,benford1938law}, population statistics and demographic data \citep[][]{raimi1976first,hill1995statistical}, and a broad class of measurements arising from multiplicative and scale-invariant processes \citep{hill1995statistical,hill1995base}. Benford's Law (henceforth, BL) predicts that the probability $P(d)$ of a digit  $ d \in \{1, 2, ..., 9\} $ appearing as the first significant digit follows a logarithmic distribution, 
\begin{equation}
    P(d) = \log_{10} \left(1 + \frac{1}{d}\right)
    \label{eq:BL}
\end{equation}

\noindent such that smaller digits occur more frequently than larger ones. This property has been widely used for anomaly detection in fields ranging from economics and forensic accounting to geophysics. In the context of seismology, \citet{sambridge2010benford} demonstrated that a variety of geophysical datasets, including seismic waveforms, exhibit conformity to BL. Subsequent studies have explored its applicability to volcanic tremor \citep{geyer2012applying}, debris-flow detection \citep{zhou2024benford}, and temporal patterns in seismicity \citep{sottili2012benford}. \citet{diaz2015ability} were among the first to investigate BL as a potential earthquake detection tool, although their analysis was limited in scope to short time windows and single-station data.
These prior studies suggest that BL statistics may capture intrinsic characteristics of seismic signals that differ from background noise. However, the potential of BL as a practical and scalable framework for local earthquake detection in continuous, long-duration, and multi-station datasets remains largely unexplored. In particular, key questions remain regarding (i) the robustness of BL-based detection across contrasting tectonic environments and noise conditions, (ii) its ability to reliably identify the onset of seismic phases, and (iii) its sensitivity to methodological choices such as sliding-window length.

In this study, we develop and evaluate a BL-based framework for local earthquake detection using continuous seismic waveform data from two contrasting geological settings in the Indian subcontinent: the intraplate Deccan Volcanic Province (DVP) and the tectonically active Himalayan region. Building on earlier works by \citet{sambridge2010benford} and \citet{diaz2015ability} we introduce a set of normalized statistical metrics that quantify deviations from background noise and emphasize transient changes in digit distributions. Using a sliding-window approach, we track the temporal evolution of BL conformity and demonstrate that local earthquake signals exhibit a systematic increase in agreement with BL distribution during the onset of seismic energy.
In this study, we address these gaps by applying a BL–based statistical framework to year‐long seismic array datasets from two contrasting geological environments in the Indian subcontinent: the intraplate Deccan Volcanic Province (DVP) and the actively deforming Himalayan region described in \textbf{Section \ref{sec:data_anal}}. We first assess whether local earthquake waveforms consistently conform to BL during the onset of seismic energy across both datasets in \textbf{Section \ref{sec:conformity}}. We then quantify the accuracy and robustness of BL‐derived statistical anomalies in identifying P‐wave onsets, and investigate the sensitivity of the method to sliding‐window length in \textbf{Section \ref{sec:detection}}.Our results demonstrate that Benford’s Law offers a computationally inexpensive, parameter‐light solution to the problem of local earthquake detection, with potential applications to large‐scale terrestrial networks as well as future planetary seismic missions.


\section{Data and Analysis}
\label{sec:data_anal}

\subsection{Data and Pre-processing}

\begin{figure}
    \includegraphics[width= \linewidth]{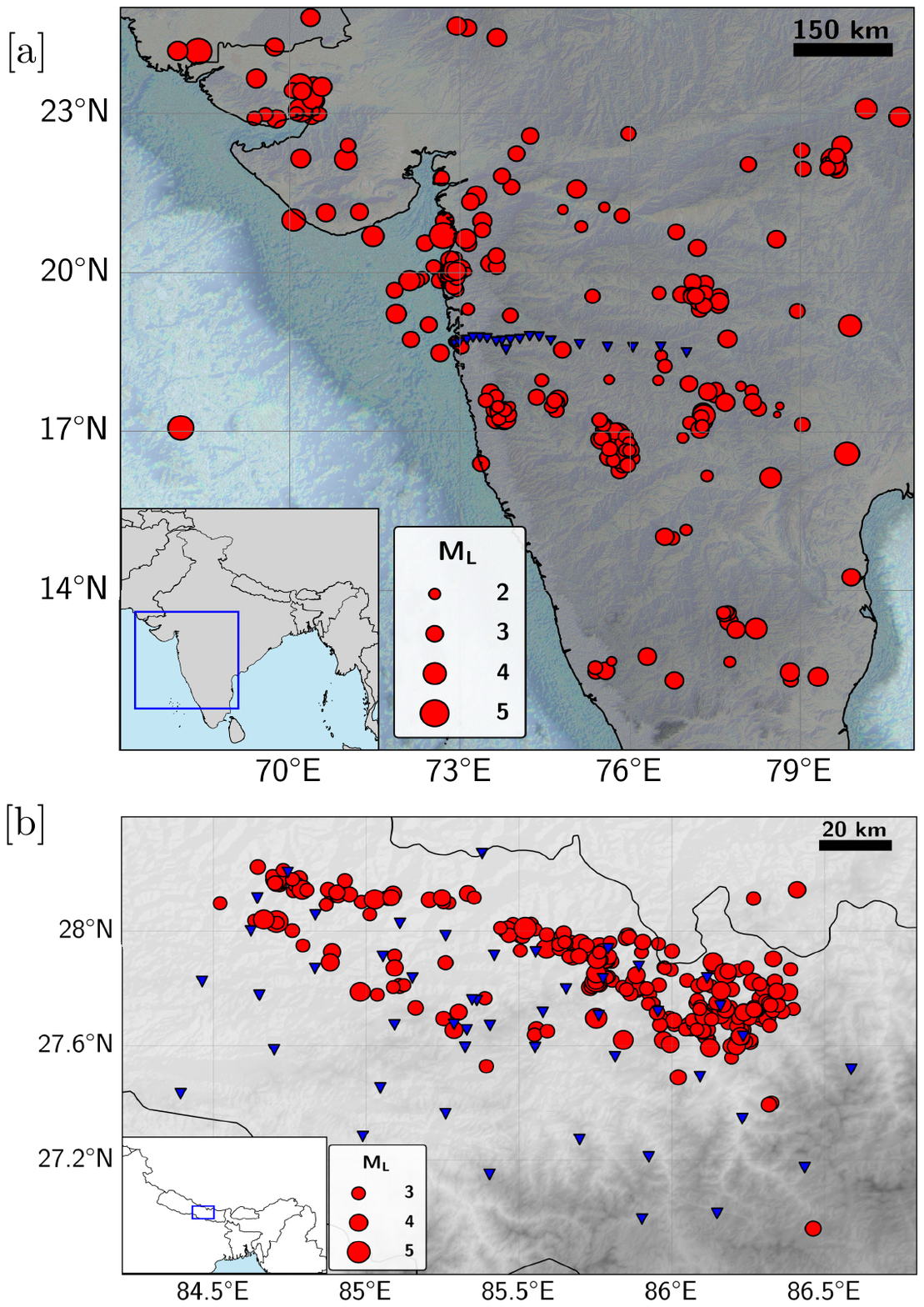}
    \caption{Station and event distribution for the datasets used in this study. (a) Map of the DVP dataset showing 20 seismic stations (triangles) and 275 events (circles). (b) Map of the NAMASTE dataset \citep{mendoza2019duplex} showing 31 seismic stations (triangles) and 250 events (circles) with $M_L > 3$.}
    
    
    \label{fig:map}
\end{figure}

To assess the robustness of our methodology, we consider two independent datasets. In each case, we analyze the continuous waveform data around known events. Our first dataset comprises broadband recordings from an array of 20 stations spanning a $\sim 430$ km long W-E profile in the Deccan Volcanic Province (DVP) of western India, deployed between January 2020 and December 2023 (Figure \ref{fig:map}a). Part of this data, corresponding to the initial phase of station deployment, was used by \cite{saha2023cryptic} to image the crust and uppermost mantle in the region. For this study, we selected earthquake events from the International Seismological Centre (ISC) catalog corresponding to the time period 1 January 2020 - 31 December 2023, and restricted to epicentral distances of less than $9^{\circ}$, thereby capturing local and near‐regional earthquakes.

The second dataset consists of aftershocks of the 2015 Gorkha earthquake in Nepal, as recorded by the 31 broadband stations of the NAMASTE network, deployed in Nepal during 2015-16 \citep{mendoza2019duplex, karplus2020rapid}. Event locations (Figure \ref{fig:map}b) were obtained from the HypoDD‐relocated catalog of \cite{mendoza2019duplex}, and we limit our study to events with $M_L>3$. This Himalayan dataset represents a tectonically active, high‐relief environment with dense aftershock sequences and elevated noise levels, providing a stringent test of the BL‐based detection framework under markedly different geological and recording conditions, compared to the DVP dataset.

Waveform data from both regions were processed identically to ensure consistency. We performed the following pre-processing:
\begin{itemize}
    \item Instrument responses were removed to obtain ground velocity in physical units (m/s), and all traces were resampled to a uniform sampling rate of 25 Hz.
    \item  A bandpass filter between 1 and 10 Hz was applied using a Hanning taper of 0.05\% to isolate frequency content typical of local earthquake signals, while minimizing spectral leakage and edge effects.
    \item For each selected event, waveform segments were extracted around the reported origin time to include a sufficiently long pre‐event noise window together with the initial portion of the seismic signal.
    \item Only the vertical component was retained for subsequent analysis, as it is most sensitive to P‐wave arrivals and provides a consistent basis for onset detection.
\end{itemize}

\subsection{Data Analysis}
Both our datasets were subjected to the following workflow, to characterize the temporal evolution of statistical properties in the seismic waveforms. 
\begin{enumerate}
    \item Each trace is divided into a series of overlapping time windows with a fixed step size of 1 s, and with window length chosen to be approximately one‐tenth of the total trace duration. This relative windowing strategy allows the analysis to adapt to variations in signal duration and noise characteristics across events and stations. While past studies have suggested fixed window lengths for specific applications \citep[e.g.,][]{diaz2015ability}, empirical testing on long‐duration array data indicates that a window length proportional to signal duration provides improved stability in noisy environments. The influence of window length on detection performance is evaluated in \textbf{Section \ref{sec:detection}}.

    \item For each time window, the first significant digit of the absolute value of each sample is extracted and used to construct a frequency distribution of leading digits. This observed distribution is then compared to the theoretical BL distribution (Eq. \ref{eq:BL}). The degree of agreement between the observed and expected digit distributions is quantified using a chi‐square misfit:

\begin{equation}
\chi^2=\sum_{d=1}^{9} \frac{(O_{d} - E_{d})^2}{E_{d}}
\label{eq:chi-square misfit}
\end{equation}

where $ O_d $ is the observed count of digit $ d $, and $ E_d $ is the expected count based on BL, calculated as $ E_d = N \cdot P(d) $, with $ N $ being the total number of samples in the window.

\item Following \cite{sambridge2010benford}, the chi‐square value is converted into a Goodness of Fit (GoF) measure:

\begin{equation}
\phi = (1 - \chi) \times 100 \%
\label{eq:GoF}
\end{equation}

The resulting GoF ($\phi$) serves as a probabilistic proxy for how well the signal follows the BL prediction, with lower $\phi$ values reflecting poor agreement with BL (suggesting noise-like behaviour) and higher $\phi$ values suggesting a good fit to BL (potentially indicating the presence of seismic events).

\item Raw GoF values are not directly comparable across stations or events due to baseline noise variability. $\phi(t)$ usually peaks ($\phi_{max}$) around the signal onset time (see Figure \ref{fig:anal1} ), but a high absolute value of $\phi_{max}$ alone is not taken as an indicator of agreement with BL.  We additionally also consider its relative rise above the baseline noise level. The baseline noise level is quantified by defining a station- and event-specific threshold GoF ($\phi_{th}$), based on the 95th percentile of GoF values from the first 200 seconds of pre-event noise. We then measure deviations from this dynamic baseline, by defining a normalized GoF deviation parameter $\Delta\phi_N(t)$:

\begin{equation}
\Delta\phi_N(t) = \frac{\phi(t)-\phi_{th}}{100-\phi_{th}}
\times\frac{\phi_{max}}{100}
\label{eq:Normalized_GoF}
\end{equation}

The parameter $\Delta\phi_N(t)$ rewards high values of $\phi_{max}$, as well as large departures from $\phi_{th}$. A high value of $\Delta\phi_N(t)$ (close to 1) occurring around the signal onset time therefore represents conformity of the signal to BL. (see Figure \ref{fig:anal1}a)

\end{enumerate}


\section{Results}

\subsection{Benford’s Law conformity during earthquakes}
\label{sec:conformity}

\begin{figure}
    \includegraphics[width=\linewidth]{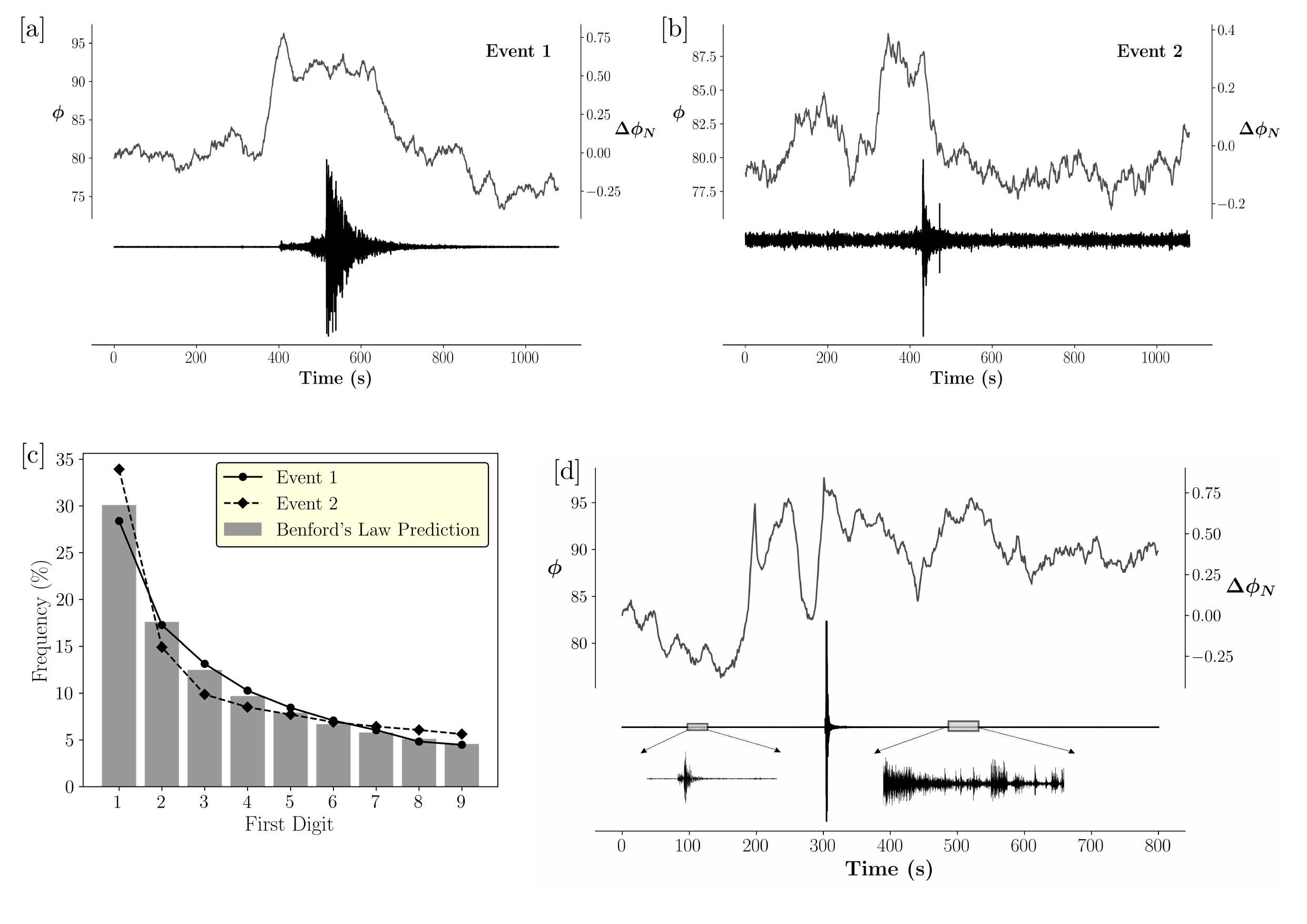}
    \caption{(a-b) Temporal variation of the Goodness of Fit ($\phi$, left axis) and the normalized deviation parameter ($\Delta \phi_{N}$, right axis) overlaid on the corresponding vertical component waveform for two events from the DVP dataset recorded at station ABR. In both cases, $\phi$ values increase sharply around the onset of seismic energy, capturing the deviation from threshold during earthquake-induced ground motion. Event 1 shows a clearer rise in $\phi(t)$ due to higher SNR compared to Event 2. The normalized deviation $\Delta \phi_{N}$ also reflects this contrast in detection sensitivity.
    (c) Comparison of first-digit distributions from the time windows corresponding to maximum $\phi$ for Event 1 and Event 2, plotted alongside the theoretical BL distribution (grey bars). Event 1 closely follows the expected logarithmic trend, while Event 2 shows greater scatter, illustrating variation in conformity to BL, based on waveform clarity and SNR.
    (d) Same analysis for a local earthquake from the NAMASTE dataset. Despite a prominent peak in $\phi$ near the P-wave onset, elevated $\phi$ values are also observed before and after the event.}
    \label{fig:anal1}
\end{figure}

\begin{figure}
    \includegraphics[width=\linewidth]{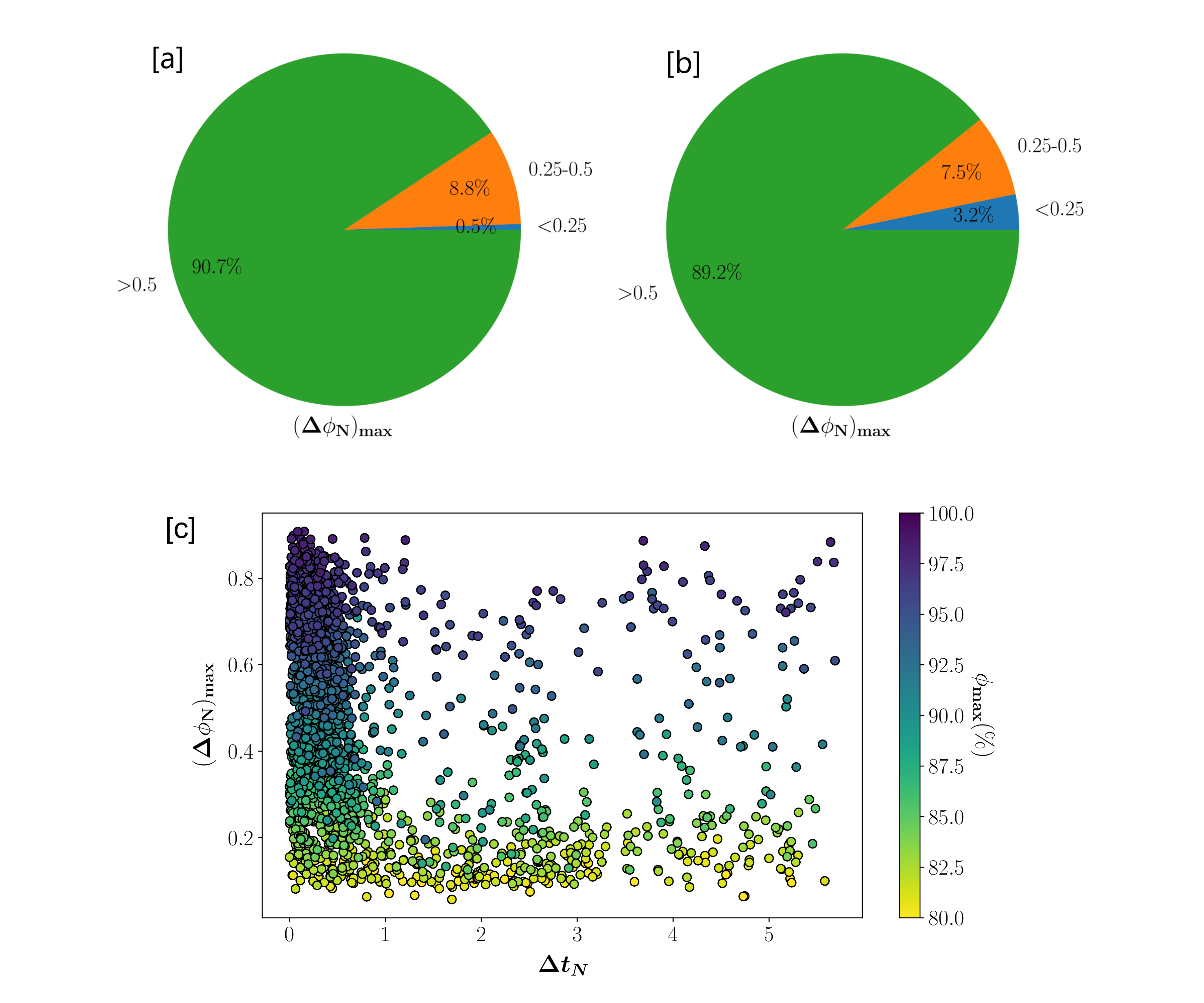}
    \caption{(a-b) Pie chart illustrating the distribution of $(\Delta\phi_N)_{max}$ across all waveforms for [a] DVP (2403 waveforms) and [b] NAMASTE (6265 waveforms) datasets. The majority of the cases exhibit $(\Delta\phi_N)_{max} > 0.5$, indicating conformity to BL around signal onset time. Only a small fraction falls below the 0.5 benchmark, reflecting weak or ambiguous detections due to poor SNR. (c) Scatter plot showing the relationship between $\Delta t_{N}$ and $\Delta \phi_{N}$ across all analyzed waveforms. Each point represents one waveform, color-coded by its maximum $\phi$ value. The clustering of points near $\Delta t_{N} < 1$ and high $\Delta \phi_{N}$ highlights the reliability of BL in identifying seismic onset with minimal lag, especially for high-quality signals.}
    \label{fig:performance}
\end{figure}

In both datasets, we first evaluate whether local earthquake signals ``conform to BL", in the sense of exhibiting a significantly enhanced agreement with BL, during the onset of seismic energy. For each waveform, we compute the Goodness of Fit ($\phi$) for all positions of the sliding window, and the corresponding normalized deviation parameter ($\Delta\phi_N$), to quantify statistically significant departures from background noise.

In the DVP dataset, $\phi$ values remain consistently low during the pre‐event noise window, reflecting poor agreement with the theoretical BL distribution. Following the onset of seismic energy, $\phi$ increases sharply, often reaching values exceeding 90\% (Figure~\ref{fig:anal1}a). This transition is consistently observed across stations and events, although the magnitude and sharpness of the increase depend on the waveform signal-to-noise ratio (SNR). Events with clearer waveforms and higher amplitudes exhibit more pronounced peaks in $\phi$ than weaker events, which show a gradual increase. The normalized deviation parameter $\Delta \phi_N$  enables a quantitative summary of our analysis. Denoting the peak value of $\Delta \phi_N(t)$ by $(\Delta \phi_N)_{max}$, we choose $(\Delta \phi_N)_{max} = 0.5$ as a reasonable cut-off value to identify BL-conformity events. For example, $\phi_{max} = 90$ yields $(\Delta \phi_N)_{max} = 0.59$ if $\phi_{th}=70$, but yields $(\Delta \phi_N)_{max} = 0.45$ if $\phi_{th}=80$. In the latter case, $\phi_{max}$ is not significantly higher than $\phi_{th}$, and our assessment parameter $(\Delta \phi_N)_{max}$ falls below the 0.5 cutoff. For easy reference, we tabulate the values of $(\Delta \phi_N)_{max}$ for some typical combinations of $\phi_{max}$ and $\phi_{th}$ observed in our data, noting that $\phi_{th}$ almost never fell below $60$ per cent.  (Table \ref{tab:delta_phi_examples}).

\begin{table}
\centering
    \begin{tabular}{ccc}
        \hline
         $\phi_{max}$& $\phi_{th}$  & $(\Delta \phi_{N})_{max}$ \\
         \hline
         95 & 85 & 0.63 \\
         \hline
         90 & 80 & 0.45 \\
         \hline
         90 & 70 & 0.59\\
         \hline
         85 & 60  & 0.53  \\
         \hline
         80 & 60 & 0.4 \\
         \hline
    \end{tabular}
    \caption{Summary of the $(\Delta \phi_{N})_{max}$ calculated for various combinations of $(\phi_{max})$ and $(\phi_{th})$ parameters.}
    \label{tab:delta_phi_examples}
\end{table}

Across the DVP dataset, the majority of waveforms (~$90.7\%$) yield $(\Delta\phi_N)_{max} > 0.5$, (Figure \ref{fig:performance}a). Only a small fraction of cases fall below this value, primarily corresponding to low SNR observations. So, we can use $(\Delta\phi_N)_{max} = 0.5$ as a benchmark for our further analysis.

A similar pattern is observed for the Himalayan dataset recorded by the NAMASTE network. Despite higher ambient noise levels and complex wave propagation effects, $\phi$ values systematically increase near the onset of seismic energy. An interesting difference, compared to the DVP dataset, is that the Himalayan waveforms show more elevated $\phi$ values beyond the P‐wave onset window, reflecting the presence of coda waves and overlapping aftershock activity (Figure~\ref{fig:anal1}d). Nevertheless, statistically significant BL conformity is established for this dataset too, through the distribution of values of $(\Delta\phi_N)_{max}$ (Figure \ref{fig:performance}b).

These results demonstrate that earthquake signals and sequence types, regardless of the specific geological environment or tectonic setting of the associated earthquakes, conform to BL inasmuch as they exhibit a much better fit to BL than pre-event noise.

\subsection{Temporal alignment with P-wave arrivals}

In the previous section we established that, relative to background noise, earthquake signals exhibit a marked improvement in the GoF to BL. However, our analysis did not place any constraints on the part of the signal that is responsible for this change. Is the rise in $\Delta\phi_N$ associated with the P-wave arrival (signal onset) or the S-wave or the surface wave train? The $(\Delta\phi_N)_{max}$ distributions presented in Figure \ref{fig:performance}(a-b) cannot answer this question, because they only consider the peak value of $\Delta\phi_N$ across a waveform, not the position at which it occurs. Intuitively, looking at Figures \ref{fig:anal1} a-b, we expect the peak to be attained at signal onset. To verify whether this is indeed the case, we use another parameter, $\Delta t_N$, that measures the temporal alignment between statistical anomalies and physical earthquake signals. First, we compute the time offset between the point of maximum GoF ($t_{\phi_{max}}$) and the P-wave arrival.

\begin{equation}
    \Delta t = t_{\phi_{max}} - t_{P}
    \label{eq: Deltat}
\end{equation}

Here, $t_{\phi_{max}}$ corresponds to the window where the maximum GoF was observed, and $t_{P}$ is a predicted P-arrival time. For large-scale testing over large number waveforms from various datasets, manual picking of P-wave onsets is not feasible. Instead, we employed the TauP Toolkit \citep{crotwell1999taup} with the AK135 global velocity model \citep{kennett1995constraints} to predict theoretical P-wave arrival times. While these predicted onsets are not exact—especially for small-magnitude local events—the prediction error is expected to be small compared to the window length. The temporal offset $\Delta t$ provides a measure of how early or late the maximum statistical anomaly occurs, relative to the physical onset of seismic arrivals from a known earthquake.
To facilitate cross-event comparison, given our variable waveform length selections (and hence variable window length), the time difference is also normalized by the sliding window length:

\begin{equation}
     \Delta t_N = \frac{\Delta t}{\text{window length}} 
     \label{eq: normalized_delta_t}
\end{equation}

The parameter $\Delta t_N$ quantifies the temporal offset between the point of maximum GoF and the approximate P-wave arrival time, in units of the sliding window length.

We only computed $\Delta t_N$ for the DVP dataset. For this dataset, event–station distances are sufficiently large that theoretical P‐wave arrival times can be robustly estimated using ray‐based travel‐time calculations. In contrast, the NAMASTE dataset consists predominantly of very close event–station pairs, for which ray paths are extremely short. Under these conditions, travel‐time predictions obtained from the TauP Toolkit become highly sensitive to near‐source velocity heterogeneity and are therefore less reliable.

The majority of waveforms exhibit $\Delta t_{N} < 1$, indicating that the maximum BL conformity occurs within one window length of the predicted P‐wave onset (Figure \ref{fig:performance}c). In many cases, the peak in $\phi$ coincides closely with the first visible increase in waveform amplitude, suggesting that BL‐based statistics are sensitive to the earliest stages of signal emergence from background noise.

\subsection{Event Detection with BL and Comparison with STA/LTA}\label{sec:compare_sta_lta}

Encouraged by the results of the previous section, we proceed to evaluate the utility of BL as an earthquake detection tool. We define a modified GoF parameter,  $\Delta\phi'_N(t)$:

\begin{equation}
\Delta\phi'_N(t) = 1 - \left[ \frac{\phi_{max}-\phi(t)}{100-\phi(t)} \right]
\label{eq:phi_prime}
\end{equation}

While $\Delta\phi_N$ effectively quantifies the overall deviation from background noise, it can remain elevated over extended portions of waveforms due to trailing aftershock activity, as was seen with the NAMASTE dataset. The parameter $\Delta\phi'_N(t)$ is designed to emphasize the gradient of $\phi$, thereby clearly highlighting the initial transition from noise-dominated to signal-dominated behavior. Importantly,  $\Delta\phi'_N(t)$ does not require any explicit threshold estimation from pre-event noise, as it is inherently normalized using $\phi(t)$ and $\phi_{max}$. This property simplifies the detection scheme and reduces sensitivity to noise-level variability across stations. As a result, $\Delta\phi'_N(t)$ improves the robustness and timing precision of BL-based P-wave onset detection, particularly in waveforms with long codas or sustained high $\phi$ values.

To place the performance of these 2 BL based metrics$\Delta\phi_N$ and $\Delta\phi'_N$ in the context of conventional detection methods, we compare their behavior with the widely used STA/LTA algorithm.

\begin{figure}
    \includegraphics[width=\linewidth]{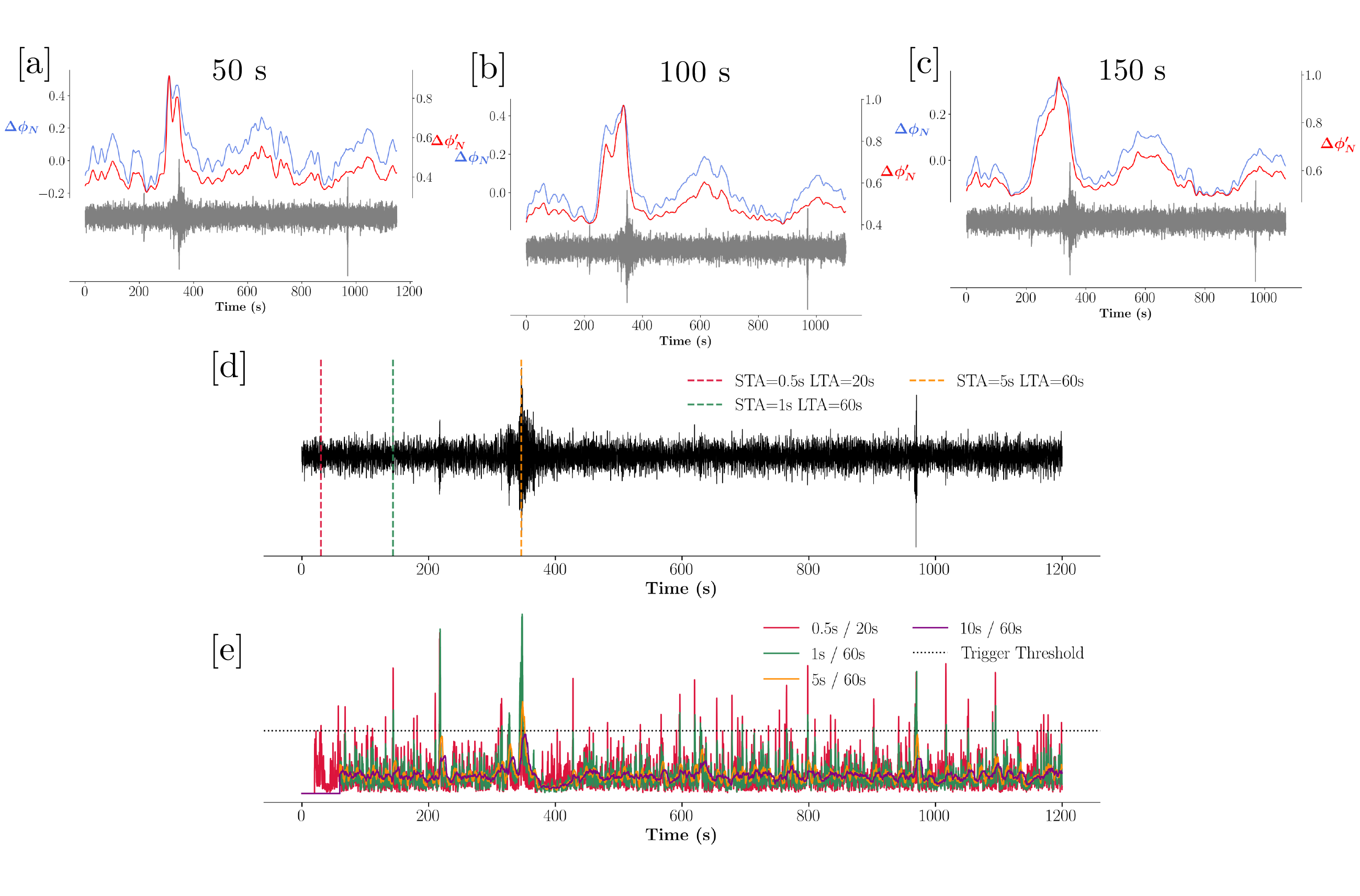}
    \caption{Comparison of BL–based detection metrics and STA/LTA performance for a representative noisy waveform. (a–c) Temporal evolution of the BL-derived parameters $\Delta\phi_N$ (blue) and $\Delta\phi'_N(t)$ (orange) computed using sliding-window lengths of 50 s, 100 s, and 150 s, respectively. (d) Same waveform with STA/LTA trigger times (vertical lines) obtained using different combinations of STA and LTA window lengths, illustrating variability in detected onset depending on parameter choice. (e) STA/LTA characteristic functions for the same parameter sets, computed using a fixed threshold value of 1.}
    \label{fig:compare_sta}
\end{figure}

Figure \ref{fig:compare_sta} illustrates this sensitivity using a representative waveform. Different combinations of STA and LTA window lengths produce markedly different characteristic functions and trigger behavior. In particular, the choice of threshold plays a crucial role: if the threshold is set too low, the method produces multiple detections, including false positives, as seen in Figure \ref{fig:compare_sta}(e). Conversely, if the threshold is set too high, valid seismic onsets may not be detected. For example, in the case of STA/LTA parameters (10 s, 60 s), the characteristic function does not exceed the chosen threshold, resulting in a missed detection despite the presence of a clear onset. This behavior highlights a fundamental limitation of STA/LTA-based detection: the need for careful and often dataset or site-specific tuning of both window lengths and threshold values to achieve reliable performance. In noisy environments, this tuning becomes particularly challenging, as parameter choices that suppress false triggers may simultaneously reduce sensitivity to actual events.

In contrast, the BL-based framework does not rely on explicit amplitude thresholds for detection. Panels (a–c) of Figure \ref{fig:compare_sta} show the temporal evolution of BL GoF for sliding-window lengths of 50 s, 100 s, and 150 s, respectively. Despite the variation in window size, $\phi$ exhibits a consistent increase around the onset of seismic energy in all cases. While the exact timing of the peak varies slightly with window length, reflecting the finite temporal resolution of the method, the detection of the event itself remains stable.

This comparison demonstrates a key distinction between the two approaches. STA/LTA requires careful tuning of both window parameters and detection thresholds, and its performance can vary significantly depending on these choices. In contrast, the BL-based framework provides a consistent and reproducible detection of seismic onset without requiring threshold adjustment, making it particularly suitable for continuous monitoring in noisy or poorly constrained environments. It is important to note that the BL-based framework identifies seismic onset as a statistical transition over a finite window, rather than a discrete sample-level discontinuity. Consequently, while detection is robust and consistent, the temporal precision of onset picking is governed by the chosen window length. 

We applied the BL-based framework to the DVP dataset using events from the ISC catalog as reference. Out of a total of 275 cataloged events within the study period and distance range, it successfully detected 268 events, corresponding to a detection rate of approximately $97.5\%$. Notably, no false positive detections were observed within the analyzed time windows. Missed detections are primarily associated with low SNR waveforms, where the statistical transition from noise to signal is less pronounced. These results demonstrate that it provides reliable event detection with high recall while maintaining a low false-trigger rate.

\subsection{P-wave onset detection Accuracy with BL}
\label{sec:detection}

we now formulate a practical detection framework based on the combined use of the BL-derived parameters $\phi(t)$, $\Delta \phi_{N} (t)$ and $\Delta \phi'_{N} (t)$ .  To identify the onset of seismic energy, we adopt a criteria-based approach in which detection is defined as the first instance where all three parameters simultaneously exceed prescribed thresholds. Specifically, we require: (i) $\phi(t) > 90 \%$, (ii)$\Delta \phi_{N} (t) > 0.5$, and (iii) $\Delta \phi'_{N} (t) > 0.5$. These thresholds are empirically determined on the basis of stability across a wide range of events and noise conditions and are chosen to ensure that detections correspond to statistically significant and temporally localized deviations from background noise. This approach ensures that detection is not triggered by isolated fluctuations in any single parameter but rather by a consistent and coherent transition across all parameters.

\begin{figure}
    \includegraphics[width=\linewidth]{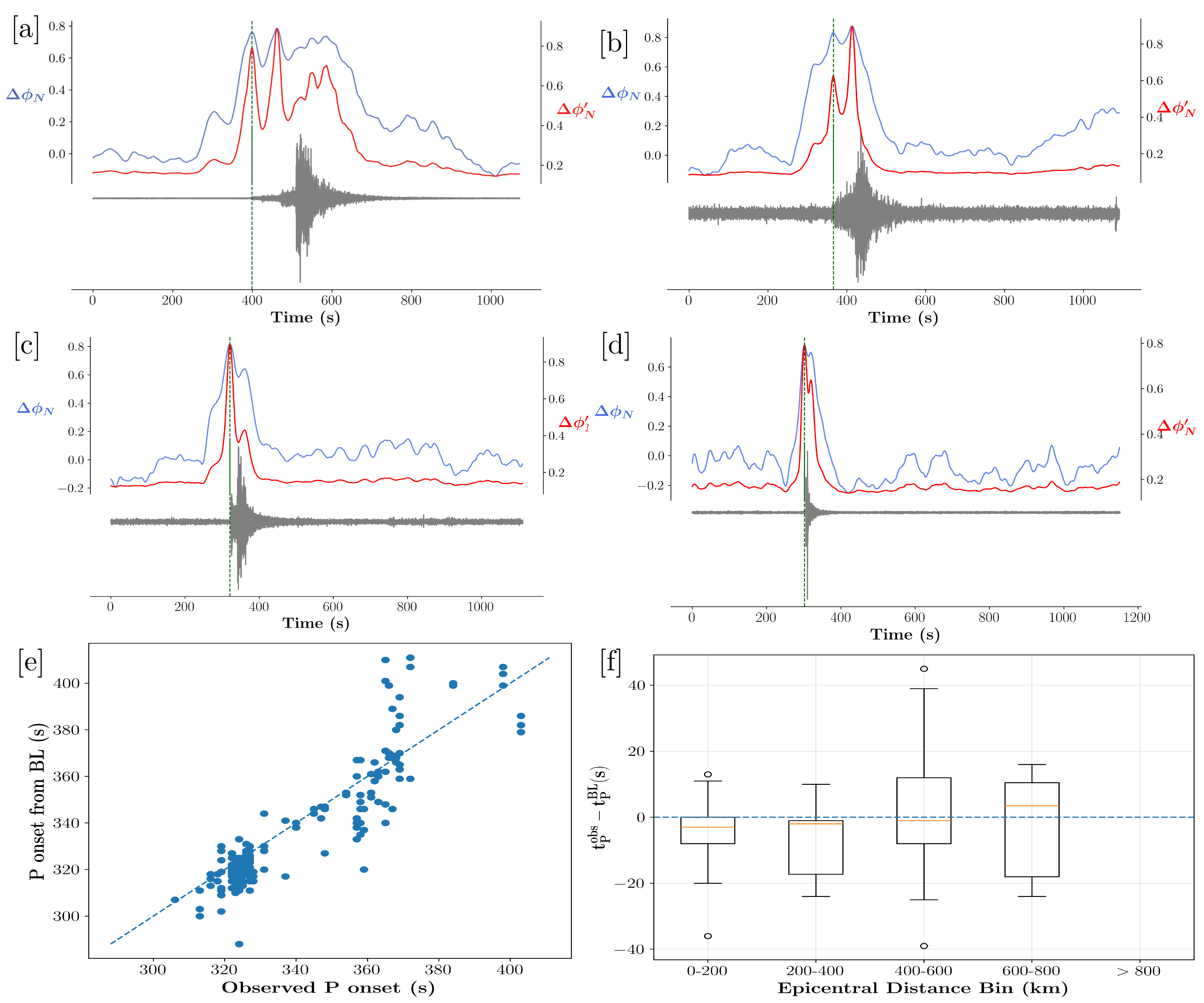}
    \caption{Performance of the normalized Benford’s Law deviation parameter $(\Delta\phi_N)$ for P-wave onset detection. (a–d) Examples of $\Delta\phi_N$ (blue) and its complementary metric $\Delta\phi'_N$ (red) overlaid on the vertical-component waveforms (grey), showing pronounced peaks near the P-wave onset (green dashed line) for representative events. (e) Comparison between observed P-wave onset times and BL-derived onsets, with the dashed line indicating the ideal one-to-one relation. (f) Residuals ($t_{P}^{obs}$ - $t_{P}^{BL}$) as a function of epicentral distance, demonstrating stable detection performance across distance bins.}

    \label{fig:best_wf_ABR}
\end{figure}

\begin{figure}
    \includegraphics[width=\linewidth]{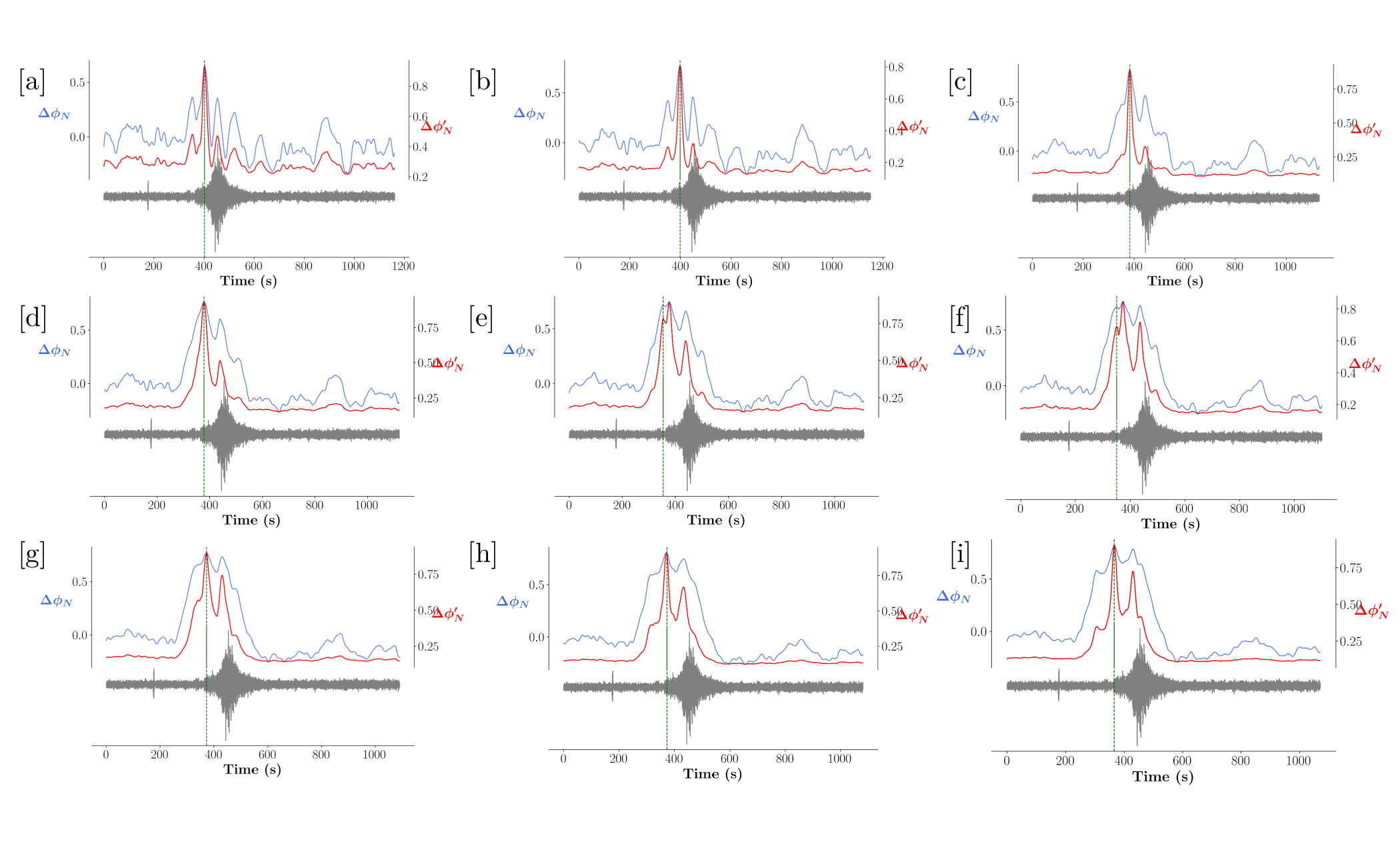}
    \caption{Sensitivity of Bl–based P onset detection to moving window length. Panels [a–i] show BL parameters ($\Delta \phi$ in blue and $\Delta \phi'$ in orange) computed for the same seismic waveform using sliding window lengths ranging from 40 s to 130 s. The seismic waveform is shown in grey, and the P onset is marked by the vertical green dashed line. The figure illustrates how window-length selection influences the performance of BL as a detection tool.}

   
    \label{fig:diff_window}
\end{figure}

\begin{figure}
    \includegraphics[width=\linewidth]{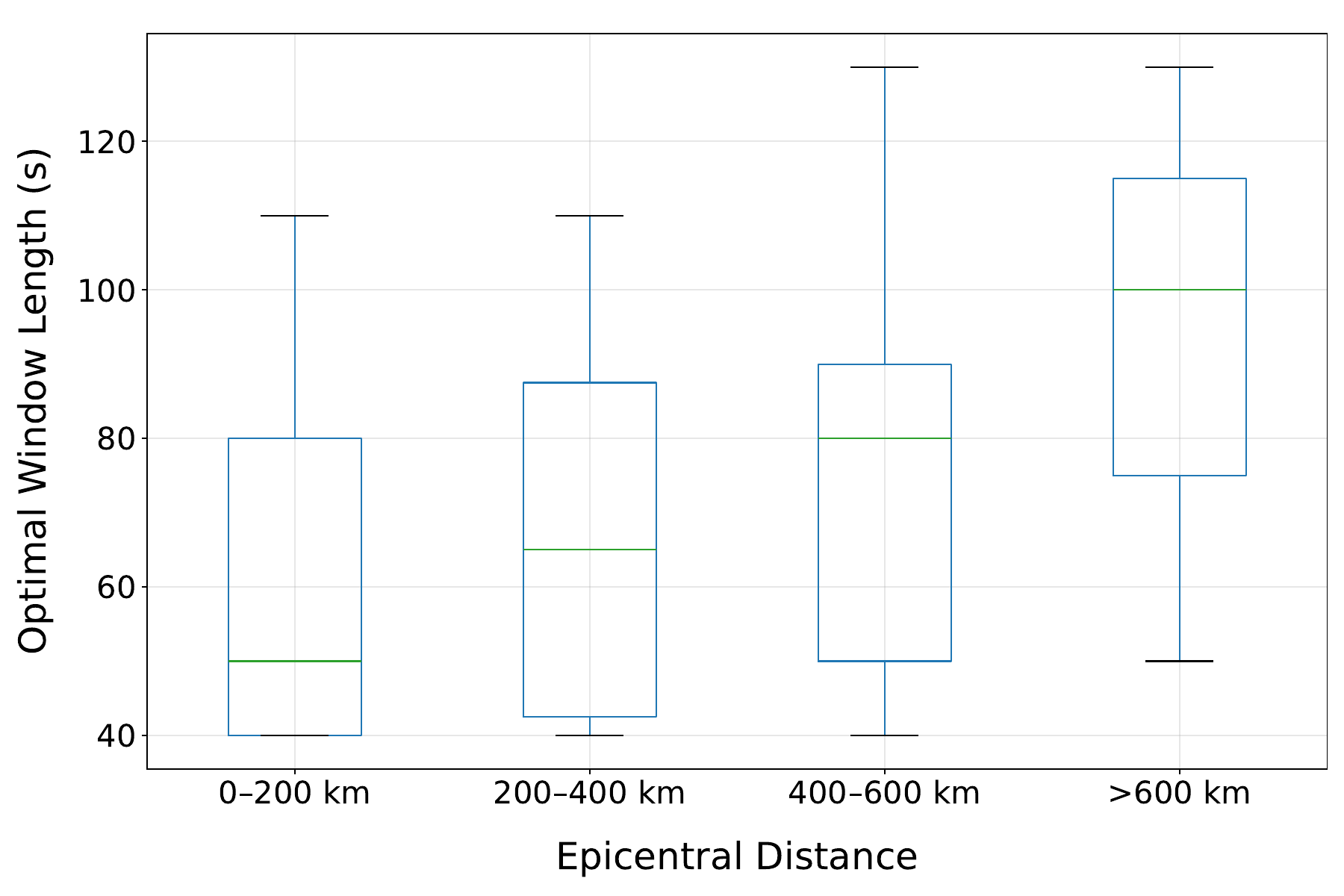}
    \caption{Variation of optimal sliding window length for BL–based P onset detection as a function of epicentral distance. Boxplot shows a systematic increase in the optimal window length with increasing distance, indicating that longer windows are required to achieve reliable BL performance for more distant events.}

   
    \label{fig:corr_window}
\end{figure}

While $\Delta \phi_{N}$ effectively quantifies the overall deviation from background noise, it can remain elevated over extended portions of waveforms due to trailing aftershock activity, as was seen with the NAMASTE dataset. The parameter $\Delta \phi'_N(t)$ is designed to emphasize the gradient of $\phi$, thereby clearly highlighting the initial transition from noise‐dominated to signal‐dominated behavior. Importantly, $ \Delta \phi'_N$ does not require any explicit threshold estimation from pre‐event noise, as it is inherently normalized using $\phi(t)$ and $\phi_{max}$. This property simplifies the detection scheme and reduces sensitivity to noise‐level variability across stations. As a result, $\Delta\phi'_N(t)$ improves the robustness and timing precision of BL‐based P‐wave onset detection, particularly in waveforms with long codas or sustained high $\phi$ values.

Armed with the three parameters $\phi(t)$, $\Delta\phi_N(t)$ and $\Delta\phi'_N(t)$, we apply a combined detection criterion requiring (i) $\phi(t) > 90\%$ , (ii) $\Delta\phi_N(t) > 0.5$ and (iii) $\Delta\phi'_N(t) > 0.5$. BL‐based onset time is defined as the first instant at which all three conditions are simultaneously satisfied for a given waveform. These criteria ensure that detections correspond to statistically significant, temporally localized deviations from background noise.

Applying this scheme to the large DVP dataset shows that P‐wave onsets can be identified robustly using BL‐based statistics alone, without explicit amplitude thresholds or phase‐specific tuning. Detection performance is strongly influenced by the choice of sliding‐window length. Shorter windows provide higher temporal resolution but are more sensitive to noise fluctuations, while longer windows yield smoother $\phi$ curves and more stable detections at the expense of timing precision.

Figure~\ref{fig:best_wf_ABR} illustrates the performance of the BL-based detector on representative waveforms for which the P-wave onset is clearly identifiable. In these examples, the rise in $\phi$ closely coincides with the visually identifiable P arrival, demonstrating that the statistical anomaly emerges contemporaneously with the onset of seismic energy. The green vertical line marks the detected onset based on BL, while the reference P-wave onset is obtained from theoretical travel-time estimates. The close alignment between these two times indicates that BL-based detection provides reliable onset timing when an appropriate window length is used.

To quantify detection accuracy across the dataset, Figure~\ref{fig:best_wf_ABR}e-f compares the BL-derived onset times with the observed P-wave arrival times. The dashed diagonal line represents the ideal case ($t_{P,obs}$ = $t_{P,BL}$), where the detected onset exactly matches the predicted arrival. The clustering of points along this line demonstrates that, for the majority of waveforms, BL-based onset times are consistent with theoretical P arrivals. Deviations from the ideal trend are primarily observed for low signal-to-noise events and recordings with emergent onsets, where the statistical transition from noise-like to signal-like behavior is more gradual.

The sensitivity of the detection to sliding-window length is illustrated in Figure~\ref{fig:diff_window}, where the same waveform is analyzed using multiple window lengths ranging from short (e.g., 40 s) to long (e.g., 130 s) durations. Shorter windows provide higher temporal resolution but produce noisier $\phi$ curves, occasionally resulting in multiple local maxima and less stable detections. As the window length increases, the time series becomes smoother, and the dominant peak aligns more consistently with the P-wave onset. However, excessively long windows can smear the onset and slightly delay the detected peak. These examples demonstrate that window length controls the trade-off between temporal precision and statistical stability.

A systematic relationship between optimal window length and epicentral distance is shown in Figure \ref{fig:corr_window}. Events recorded at larger epicentral distances, which generally exhibit longer-duration and more emergent waveforms, require larger window lengths to achieve stable and accurate detections. In contrast, nearby events with impulsive P onsets are best resolved using shorter windows. This trend indicates that window length should scale with the characteristic duration of the signal, rather than being fixed across all events.


These observations indicate that the window length defines the intrinsic temporal resolution of the BL-based method, introducing a controlled trade-off between statistical stability and timing precision. Importantly, this parameter affects only the precision of onset timing, while the detection of the event itself remains robust across a wide range of window lengths, as demonstrated in \textbf{Section \ref{sec:compare_sta_lta}}.


\section{Discussion and Conclusion}

This study demonstrates that BL provides a robust statistical framework for detecting local earthquake signals in continuous seismic data. Unlike conventional energy-based detectors such as STA/LTA, which rely on amplitude ratios and require careful tuning of window lengths and threshold parameters, the BL-based approach identifies seismic onset as a transition in the statistical distribution of waveform amplitudes. This fundamental difference leads to a detection framework that is inherently stable and largely insensitive to parameter selection.

At the same time, the BL-based method should not be interpreted as a high-resolution phase picker. Because detection is based on statistical behavior over a finite window, the method identifies the onset of seismic energy as a temporal interval rather than a single sample-level discontinuity. Consequently, the temporal precision of phase picking is governed by the chosen window length, which controls the trade-off between statistical stability and timing resolution. Shorter windows improve temporal precision but are more sensitive to noise, whereas longer windows provide more stable detection at the expense of timing accuracy.
Within this framework, the BL-based approach is best viewed as a robust, parameter-light detector rather than a replacement for advanced picking algorithms. Its primary strength lies in its ability to consistently identify the transition from noise to signal without requiring prior training, waveform templates, or extensive parameter optimization. This makes it particularly suitable for deployment in data-sparse environments, noisy conditions, and large-scale continuous monitoring systems.

A key implication of this work is the potential role of BL-based detection within multi-stage processing pipelines. Rather than competing directly with machine learning (ML) or deep learning (DL) based pickers, the BL-based method can serve as an efficient pre-triggering or pre-filtering mechanism, identifying candidate time windows that are likely to contain seismic events. These candidate segments can then be passed to more computationally intensive ML/DL models for precise phase picking and classification. Such a hybrid approach offers a practical balance between computational efficiency and detection performance, reducing the overall processing load while retaining sensitivity to seismic signals.

This capability is particularly relevant for real-time monitoring systems and large datasets, where applying advanced ML/DL models continuously to entire waveform records may be computationally expensive. By filtering out noise-dominated segments and focusing analysis on statistically significant intervals, the BL-based framework can significantly reduce the volume of data requiring detailed processing.

Despite these advantages, some limitations remain. The method depends on the choice of sliding-window length, which determines the temporal resolution of detection. While this parameter is less critical than the multiple parameters required in STA/LTA-based methods, it still introduces a trade-off between precision and stability. Additionally, performance may be reduced for extremely weak or highly emergent signals, where the statistical transition from noise to signal is less distinct.

In summary, this study demonstrates that Benford’s Law provides a computationally efficient, parameter-lite solution for local earthquake detection. By introducing complementary statistical metrics and validating them across the contrasting Deccan and Himalayan datasets, we show that BL-conformity is a universal property of seismic onset. This framework offers a scalable alternative for terrestrial network monitoring, and it could also be a promising tool for future planetary seismic missions, where data availability and prior geological constraints are inherently limited.


\bibliographystyle{plainnat}
\bibliography{benford}

@article{raimi1976first,
  title={The first digit problem},
  author={Raimi, Ralph A},
  journal={The American Mathematical Monthly},
  volume={83},
  number={7},
  pages={521--538},
  year={1976},
  publisher={Taylor \& Francis}}

@article{hill1995base,
  title={Base-invariance implies Benford’s law},
  author={Hill, Theodore P},
  journal={Proceedings of the American Mathematical Society},
  volume={123},
  number={3},
  pages={887--895},
  year={1995}}

@article{hill1995statistical,
  title={A statistical derivation of the significant-digit law},
  author={Hill, Theodore P},
  journal={Statistical science},
  pages={354--363},
  year={1995},
  publisher={JSTOR}}

@article{newcomb1881note,
  title={Note on the frequency of use of the different digits in natural numbers},
  author={Newcomb, Simon},
  journal={American Journal of mathematics},
  volume={4},
  number={1},
  pages={39--40},
  year={1881},
  publisher={JSTOR}}

@article{li2024saipy,
  title={SAIPy: A Python package for single-station earthquake monitoring using deep learning},
  author={Li, Wei and Chakraborty, Megha and Cartaya, Claudia Quinteros and K{\"o}hler, Jonas and Faber, Johannes and Meier, Men-Andrin and R{\"u}mpker, Georg and Srivastava, Nishtha},
  journal={Computers \& Geosciences},
  volume={192},
  pages={105686},
  year={2024},
  publisher={Elsevier}}

@article{suarez2025picking,
  title={Picking regional seismic phase arrival times with deep learning},
  author={Suarez, Albert Leonardo Aguilar and Beroza, Gregory},
  journal={Seismica},
  volume={4},
  number={1},
  year={2025}}

@article{karplus2020rapid,
  title={{A rapid response network to record aftershocks of the 2015 M 7.8 Gorkha earthquake in Nepal}},
  author={Karplus, Marianne S and Pant, Mohan and Sapkota, Soma Nath and N{\'a}b{\v{e}}lek, John and Velasco, Aaron A and Adhikari, Lok Bijaya and Ghosh, Abhijit and Klemperer, Simon L and Kuna, Vaclav and Mendoza, Manuel M and others},
  journal={Seismological Research Letters},
  volume={91},
  number={4},
  pages={2399--2408},
  year={2020},
  publisher={Seismological Society of America}}

@article{zhou2025AI-PAL,
author = {Zhou, Yijian and Ding, Hongyang and Ghosh, Abhijit and Ge, Zengxi},
title = {{AI-PAL: Self-Supervised AI Phase Picking via Rule-Based Algorithm for Generalized Earthquake Detection}},
journal = {Journal of Geophysical Research: Solid Earth},
volume = {130},
number = {4},
pages = {e2025JB031294},
doi = {https://doi.org/10.1029/2025JB031294},
url = {https://agupubs.onlinelibrary.wiley.com/doi/abs/10.1029/2025JB031294},
year = {2025}}

@article{zhou2021an,
    author = {Zhou, Yijian and Yue, Han and Fang, Lihua and Zhou, Shiyong and Zhao, Li and Ghosh, Abhijit},
    title = {{An Earthquake Detection and Location Architecture for Continuous Seismograms: Phase Picking, Association, Location, and Matched Filter (PALM)}},
    journal = {Seismological Research Letters},
    volume = {93},
    number = {1},
    pages = {413-425},
    year = {2021},
    month = {09},
    doi = {10.1785/0220210111},
    url = {https://doi.org/10.1785/0220210111}}

@article{Allen1978automatic,
    author = {Allen, Rex V.},
    title = {{Automatic earthquake recognition and timing from single traces}},
    journal = {Bulletin of the Seismological Society of America},
    volume = {68},
    number = {5},
    pages = {1521-1532},
    year = {1978},
    month = {10},
    issn = {0037-1106},
    doi = {10.1785/BSSA0680051521},
    url = {https://doi.org/10.1785/BSSA0680051521},
    eprint = {https://pubs.geoscienceworld.org/ssa/bssa/article-pdf/68/5/1521/5321384/bssa0680051521.pdf}}

@article{zhou2024benford,
  title={{Benford's law as debris flow detector in seismic signals}},
  author={Zhou, Qi and Tang, Hui and Turowski, Jens M and Braun, Jean and Dietze, Michael and Walter, Fabian and Yang, Ci-Jian and Lagarde, Sophie},
  journal={Journal of Geophysical Research: Earth Surface},
  volume={129},
  number={9},
  pages={e2024JF007691},
  year={2024},
  publisher={Wiley Online Library}}

@article{geyer2012applying,
  title={{Applying Benford's law to volcanology}},
  author={Geyer, Adelina and Mart{\'\i}, Joan},
  journal={Geology},
  volume={40},
  number={4},
  pages={327--330},
  year={2012},
  publisher={Geological Society of America}}

@article{kennett1995constraints,
  title={{Constraints on seismic velocities in the Earth from traveltimes}},
  author={Kennett, Brian LN and Engdahl, ER and Buland, Raymond},
  journal={Geophysical Journal International},
  volume={122},
  number={1},
  pages={108--124},
  year={1995},
  publisher={Blackwell Publishing Ltd Oxford, UK}}

@article{crotwell1999taup,
  title={The TauP Toolkit: Flexible seismic travel-time and ray-path utilities},
  author={Crotwell, H Philip and Owens, Thomas J and Ritsema, Jeroen and others},
  journal={Seismological Research Letters},
  volume={70},
  pages={154--160},
  year={1999},
  publisher={Seismological Society of America}}

@article{sottili2012benford,
  title={Benford’s Law in time series analysis of seismic clusters},
  author={Sottili, Gianluca and Palladino, Danilo M and Giaccio, Biagio and Messina, Paolo},
  journal={Mathematical Geosciences},
  volume={44},
  number={5},
  pages={619--634},
  year={2012},
  publisher={Springer}}

@article{mendoza2019duplex,
  title={{Duplex in the Main Himalayan Thrust illuminated by aftershocks of the 2015 M w 7.8 Gorkha earthquake}},
  author={Mendoza, MM and Ghosh, A and Karplus, MS and Klemperer, SL and Sapkota, SN and Adhikari, LB and Velasco, A},
  journal={Nature Geoscience},
  volume={12},
  number={12},
  pages={1018--1022},
  year={2019},
  publisher={Nature Publishing Group UK London}}

@article{sambridge2010benford,
  title={{Benford}'s law in the natural sciences},
  author={Sambridge, Malcolm and Tkal{\v{c}}i{\'c}, Hrvoje and Jackson, A},
  journal={Geophysical research letters},
  volume={37},
  number={22},
  year={2010},
  publisher={Wiley Online Library}
}

@article{diaz2015ability,
  title={On the ability of the {Benford}’s law to detect earthquakes and discriminate seismic signals},
  author={D{\'\i}az, J and Gallart, J and Ruiz, M},
  journal={Seismological Research Letters},
  volume={86},
  number={1},
  pages={192--201},
  year={2015},
  publisher={Seismological Society of America}}

@article{saha2023cryptic,
  title={{Cryptic magma chamber in the Deccan Traps imaged using receiver functions and surface wave dispersion}},
  author={Saha, Gokul and Kumar, Vivek and Chaubey, Dipak K and Rai, Shyam S},
  journal={Geophysical Research Letters},
  volume={50},
  number={23},
  pages={e2023GL105359},
  year={2023},
  publisher={Wiley Online Library}}

@article{benford1938law,
  title={The law of anomalous numbers},
  author={Benford, Frank},
  journal={Proceedings of the American Philosophical Society},
  pages={551--572},
  year={1938},
  publisher={JSTOR}}

\end{document}